\documentclass[a4paper,nofootinbib,superscriptaddress]{revtex4}

\usepackage{color}
\usepackage{amsmath,amsfonts,amssymb,amsthm}
\usepackage{graphicx}
\usepackage{fullpage}
\usepackage{subcaption}
\usepackage{enumerate}

\usepackage{epstopdf} 
\usepackage{bm}  

\usepackage[pdfpagelabels,pdftex,bookmarks,breaklinks]{hyperref}
\definecolor{darkblue}{RGB}{0,0,127} 
\definecolor{darkgreen}{RGB}{0,150,0}
\hypersetup{colorlinks, linkcolor=darkblue, citecolor=darkgreen, filecolor=red, urlcolor=blue}

\usepackage{layout}

\newtheorem{theorem}{Theorem}

\def\Z{\mathbb{Z}}
\def\R{\mathbb{R}}

\def\F{\mathbb{F}}

\def\e{\mathrm{e}}

\newcommand{\Eref}[1]{Eq.~(\ref{#1})}
\newcommand{\Sref}[1]{Sec.~\ref{#1}}
\newcommand{\Fref}[1]{Fig.~\ref{#1}}

\newcommand{\Tref}[1]{Thm.~\ref{#1}}

\DeclareMathOperator{\im}{im}

\newcommand{\sizedexpect}[1]{\left\langle{#1}\right\rangle}

\begin{document}

\title{A proposal for self-correcting stabilizer quantum memories in 3 dimensions (or slightly less)}

\author{Courtney G.\ Brell}
\email{courtney.brell@gmail.com}
\affiliation{Institut f\"ur Theoretische Physik, Leibniz Universit\"at Hannover, Appelstra\ss e 2, 30167 Hannover, Germany}


\begin{abstract}
We propose a family of local CSS stabilizer codes as possible candidates for self-correcting quantum memories in 3D. The construction is inspired by the classical Ising model on a Sierpinski carpet fractal, which acts as a classical self-correcting memory. Our models are naturally defined on fractal subsets of a 4D hypercubic lattice with Hausdorff dimension less than 3. Though this does not imply that these models can be realised with local interactions in $\R^3$, we also discuss this possibility. The $X$ and $Z$ sectors of the code are dual to one another, and we show that there exists a finite temperature phase transition associated with each of these sectors, providing evidence that the system may robustly store quantum information at finite temperature.
\end{abstract}

\maketitle


\newcommand{\dimh}{\mathrm{dim}}

\section{Introduction}

	There is significant interest from both an abstract and practical perspective as to if and how self-correcting quantum memories might be realised. A practical self-correcting memory would allow for arbitrarily long storage of quantum information at finite temperature without the need for constant active error-correction techniques. The 4D toric code is a simple, exactly solvable example of a system with local interactions in 4 spatial dimensions that is known to have self-correcting properties~\cite{Dennis2002, Alicki2010}. In 2D, the toric code is known to be unstable at finite temperature~\cite{Nussinov2008, Nussinov2009, Alicki2009}, and there are numerous no-go theorems that rule out broad classes of models for self-correction~\cite{Kay2008,Bravyi2009,Nussinov2012, Landon-Cardinal2013}. Despite this, some attempts have been made to engineer self-correcting behaviour in 2D systems~\cite{Hamma2009, Brown2013, Kapit2014}. Many approaches towards realising some aspects of self-correction in 3D have also been found, notably including the Haah code~\cite{Haah2011,Bravyi2011a,Bravyi2011b, Bravyi2013} among others~\cite{Bacon2006, Chesi2010, Pastawski2011, Hutter2012, Michnicki2012, Becker2013, Pedrocchi2013}, though no local spin models in 2D or 3D are known to be fully self-correcting. There are also several no-go results restricting possible self-correcting models in 3D~\cite{Pastawski2010, Yoshida2011, Nussinov2012, Haah2013, Pastawski2014}. For a comprehensive review of quantum memories at finite temperature, see Ref.~\cite{BrownReview}.
	
	We propose here a local spin model with dimension less than 3, and argue that it may act as a self-correcting quantum memory. Our approach is based on fractal geometries, and inspired by the classical self-correcting behaviour of an Ising model on a Sierpinski carpet graph. The Sierpinski carpets~\cite{Sierpinski1916} are a family of fractal subsets of $\R^2$ with Hausdorff dimension between 1 and 2. We propose a family of quantum CSS codes that can be considered as 4D toric codes on discretizations of the product of two Sierpinski carpet fractals (with appropriate boundary conditions). Concretely, our codes are defined through the homological product construction~\cite{Freedman2014, Bravyi2013homological} applied to two toric codes on 2D Sierpinski carpet graphs, yielding a code family with extensive degeneracy. Though these models naturally embed in $\R^4$, their Hausdorff dimension may be chosen to be less than 3. We call these codes \emph{fractal product codes} (FPCs). We also discuss the prospect of realizing these codes locally in $\R^3$, though we do not prove that this is possible.
	
	Though we call them codes, FPCs should more properly be considered Hamiltonian systems given by a (negative) sum of stabilizer generators, and we show that such systems have (at least) two phase transitions at finite temperature, one associated with each sector ($X$ or $Z$) of the CSS code. The tools we use to show this are generalized duality transformations and correlation inequalities. Given these phase transitions, we argue that the FPC system may act as a self-correcting quantum memory at sufficiently low temperatures. Though there is an extensive degeneracy, we expect the phase transitions we identify to correspond to the appearance of thermal stability for one preferred encoded qubit associated with global degrees of freedom.
	
	The use of the Sierpinski carpet fractals is not crucial for our construction, and so we also briefly discuss the more general family of product codes that could arise from other graphs, such as alternative fractal graphs.

	\subsection{The Caltech rules}
		A practical quantum memory could in principle take any number of forms. In order to concretely discuss a self-correcting quantum memory (SCQM), it is convenient to set a series of criteria which such a system should satisfy. As such, we briefly review the so-called \emph{Caltech rules}:
	
		A model is a $D$-dimensional SCQM under the Caltech rules if:
		\begin{enumerate}
			\item (finite spins) It consists of finite dimensional spins embedded in $\R^D$ with finite density
			\item (bounded local interactions) It evolves under a Hamiltonian comprised of a finite density of interactions of bounded strength and bounded range
			\item (nontrivial codespace) It encodes at least one qubit in its degenerate ground space
			\item (perturbative stability) The logical space associated with at least one encoded qubit must be perturbatively stable in the thermodynamic limit
			\item (efficient decoding) This encoded qubit allows for a polynomial time decoding algorithm
			\item (exponential lifetime) Under coupling to a thermal bath at some non-zero temperature in the weak-coupling Markovian limit, the lifetime of this encoded qubit asymptotically scales exponentially in the number of spins
		\end{enumerate}
	
		We purposely leave the precise definition of perturbative stability vague, and will discuss it further in \Sref{s:pertstab}. It is also often required that the Hamiltonian be gapped, but while this may be desirable it is not a necessary condition for self-correcting behaviour.
		
		The 4D toric code is an example of a 4D Caltech SCQM. Other proposals in 2 or 3 dimension often make use of long-range interactions, bosonic modes in place of spins, or do not achieve asymptotically exponential memory lifetime. Depending on the context, such relaxations of the Caltech rules may be perfectly reasonable strategies to produce a practical passive quantum information storage device. The rules as stated above are largely inspired by analogy to classically self-correcting Ising models, and are by design extremely restrictive. Alternative criteria for self-correcting memories are also discussed in~\cite{BrownReview}.
		
		Although we believe our proposal may be a candidate for a 3D Caltech SCQM, we stress that we merely provide suggestive arguments, and do not prove, that it satisfies the majority of these constraints.
		

\section{Fractals and dimensionality}

	Key to our construction will be the notion of fractal geometry.	Fractal objects have spatial dimension that interpolates between the familiar integral topological dimensions. This dimension can be quantified in several useful ways, such as the Hausdorff dimension or the box-counting dimension. We will not give details of many results familiar in fractal geometry, instead we refer the interested reader to a standard text such as Ref.~\cite{FalconerBook}.	
	
	We will consider only fractals that are particularly well-behaved, in that they are self-similar Borel sets satisfying the open set condition. For fractals with these properties, many different fractal dimensions coincide, and so we will simply denote the dimension of a fractal $F$ as $\dimh F$ (for concreteness this can be taken as the Hausdorff dimension). The dimensions of these sets also satisfy $\dimh (F_1\times F_2)=\dimh F_1+\dimh F_2$, which will be a useful identity.

	\subsection{Sierpinski carpets}
	
		Our construction is motivated by a particular family of fractals, the Sierpinski carpets~\cite{Sierpinski1916}. These fractals have dimension between 1 and 2, and are naturally defined as self-similar subsets of $\R^2$. Although more general definitions of Sierpinski carpets are sometimes used, for our purposes it will be sufficient to define a Sierpinski carpet by two positive integers $b$ and $c$, with $(b-c)$ even and positive (following e.g.~\cite{Monceau1998} or \cite{Bonnier1987}). We denote the resulting fractals by $SC(b,c)$.
		
\begin{figure}
			\begin{subfigure}{0.19\textwidth}
			\includegraphics{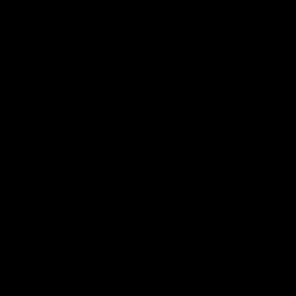}
			\end{subfigure}
			\begin{subfigure}{0.19\textwidth}
			\includegraphics{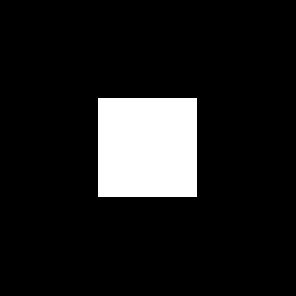}
			\end{subfigure}
			\begin{subfigure}{0.19\textwidth}
			\includegraphics{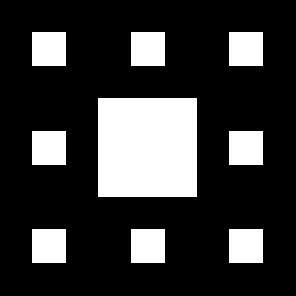}
			\end{subfigure}
			\begin{subfigure}{0.19\textwidth}
			\includegraphics{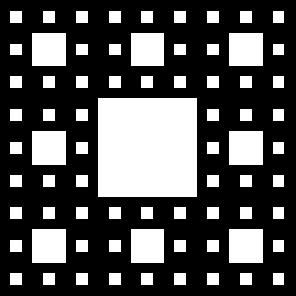}
			\end{subfigure}
			\begin{subfigure}{0.19\textwidth}
			\includegraphics{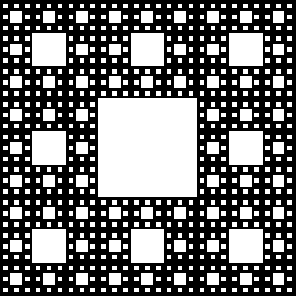}
			\end{subfigure}
			\caption{$SC(3,1,l)$ for $l=0$ to $4$}\label{f:sierpinskilevels}
		\end{figure}

		The fractals $SC(b,c)$ can be defined as the limit of a sequence $SC(b,c,l)$ as $l\to\infty$. We call the $SC(b,c,l)$ the $(b,c)$ Sierpinski carpets at level $l$, and they are constructed by dividing the unit square into $b^2$ smaller squares, deleting the central $c^2$ squares, and iterating this procedure $l$ times. An example is shown in \Fref{f:sierpinskilevels}. The dimension of a $(b,c)$ Sierpinski carpet is $\dimh SC(b,c)=\frac{\ln(b^2-c^2)}{\ln(b)}$. For positive integral $b$, $c$, and $(b-c)$, it is clear that achievable dimensions are dense in the interval $1<\dimh SC(b,c)< 2$ (and empty outside).
		
	\subsection{Sierpinski carpet graphs}		
		
\begin{figure}
			\begin{subfigure}{0.19\textwidth}
			\includegraphics{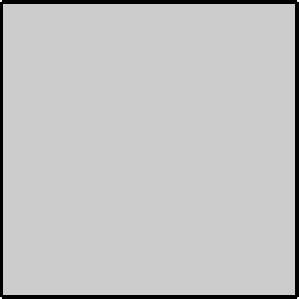}
			\end{subfigure}
			\begin{subfigure}{0.19\textwidth}
			\includegraphics{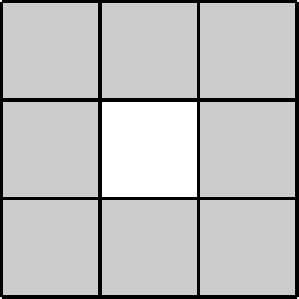}
			\end{subfigure}
			\begin{subfigure}{0.19\textwidth}
			\includegraphics{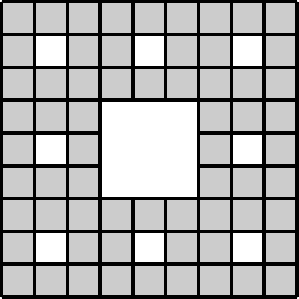}
			\end{subfigure}
			\begin{subfigure}{0.19\textwidth}
			\includegraphics{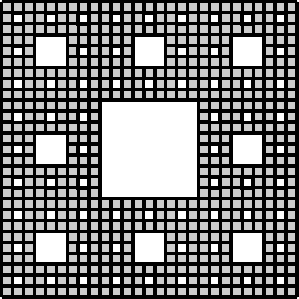}
			\end{subfigure}
			\begin{subfigure}{0.19\textwidth}
			\includegraphics{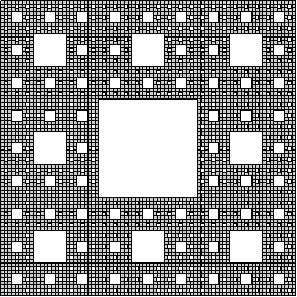}
			\end{subfigure}
			\caption{$\widehat{SC}(3,1,l)$ for $l=0$ to $4$, overlaid on $SC(3,1,l)$}\label{f:sierpinskigraph}
		\end{figure}

		At level $l$, a $(b,c)$ Sierpinski carpet is an arrangement of $(b^2-c^2)^l$ out of a possible $b^{2l}$ elementary squares. Drawing the borders of these squares yields the Sierpinski carpet graphs, i.e.~with a vertex at each corner of occupied squares (note alternative conventions exist in the literature, see~\cite{Monceau1998} or \cite{Bonnier1987} for discussion). We will denote such graphs by $\widehat{SC}(b,c,l)$. Examples are shown in \Fref{f:sierpinskigraph}. These graphs consist of
		\begin{align}
		|V(b,c,l)|&=(b^2-c^2)^l+2c\frac{b^l-(b^2-c^2)^l}{b-(b^2-c^2)}-\frac{1-(b^2-c^2)^l}{1-(b^2-c^2)}+2b^l+1\;\;\;\mbox{vertices, and}\\
		|E(b,c,l)|&=2(b^2-c^2)^l+2c\frac{b^l-(b^2-c^2)^l}{b-(b^2-c^2)}+2b^l\;\;\;\mbox{edges.}
		\end{align}
		
		It will be convenient to distinguish between ``interior'' and ``exterior'' plaquettes of the Sierpinski carpet graphs. The interior plaquettes are those bounding occupied squares of the Sierpinski carpet fractal, while the exterior plaquettes are the minimal cycles not generated by interior ones (i.e.~those cycles bounding ``deleted'' regions of the fractal, plus the outer boundary). $\widehat{SC}(b,c,l)$ can be shown to contain $|P_i(b,c,l)|$ interior plaquettes and $|P_e(b,c,l)|$ independent exterior plaquettes, with
		\begin{align}
			|P_i(b,c,l)|&=(b^2-c^2)^l\;\;\; \mbox{and} \\
			|P_e(b,c,l)|&=\frac{1-(b^2-c^2)^l}{1-(b^2-c^2)}\ ,
		\end{align}
		noting that the outer boundary can be generated by the product of all interior and exterior plaquettes.
		
		We will sometimes use the term ``Sierpinski carpet'' to refer to either the fractal or graph when it is clear which object is meant from context.
		
\subsection{Sierpinski carpet Ising models}\label{s:scising}
		
\begin{figure}
			\begin{subfigure}{0.23\textwidth}
			\includegraphics{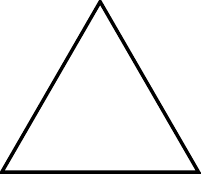}
			\end{subfigure}
			\begin{subfigure}{0.23\textwidth}
			\includegraphics{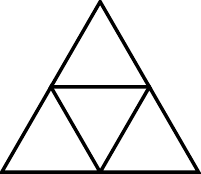}
			\end{subfigure}
			\begin{subfigure}{0.23\textwidth}
			\includegraphics{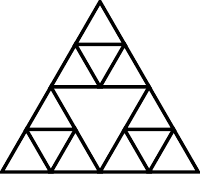}
			\end{subfigure}
			\begin{subfigure}{0.23\textwidth}
			\includegraphics{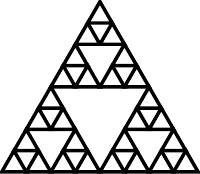}
			\end{subfigure}
			\caption{Sierpinski triangle graphs at levels $1$ to $4$}\label{f:sierpinskitrigraph}
		\end{figure}

		It is possible to define a classical ferromagnetic Ising model on a Sierpinski carpet graph, and study the thermodynamic properties for fixed $b$ and $c$ as $l\to\infty$. These models have 2-fold degenerate ground spaces, and thus can be considered as classical codes. General arguments suggest~\cite{Gefen1980, Gefen1984III}, many numerical studies demonstrate (e.g.~\cite{Monceau1998,Bonnier1987}), and it can be rigorously proved~\cite{Shinoda2002, Vezzani2003, Campari2010}, that such a family of Ising models has a phase transition at non-zero temperature. This behaviour is similar to the 2D Ising model, but in contrast to the 1D Ising model which famously has no ordered phase at finite temperature. Intuitively, the existence of the phase transition is due to the fact that the Sierpinski carpet graphs have infinite ramification order (i.e.~in the limit $l\to \infty$, an infinite number of bonds must be cut to separate the graph into two infinite pieces). General arguments suggest that an Ising model defined on any family of fractal graphs with sufficiently large ramification (in our context, scaling fast enough with $l$) will have a finite temperature phase transition, while those with finite ramification order do not exhibit finite-temperature phase transitions. The Sierpinski triangle graphs (\Fref{f:sierpinskitrigraph}) are an example of a family of fractal graphs with finite ramification, and the corresponding Ising model has only a zero-temperature phase transition~\cite{Gefen1984II}, much like the 1D Ising model.

\begin{figure}
			\includegraphics{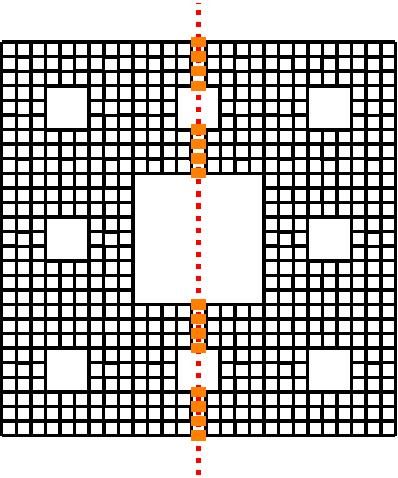}
			\caption{A minimal cut through $\widehat{SC}(3,1,3)$ shown in red, with frustrated bonds highlighted in orange.}\label{f:sierpinskigraphcut}
		\end{figure}

		Another quantity of interest is the minimum energy barrier that must be overcome to transition between the two degenerate code states by a sequence of spin flips. While this quantity is not as important as the existence of a phase transition, we can nonetheless compute it. In order to transition between the two code states, every spin on the lattice must be flipped. The energy barrier must then be at least proportional to the number of bonds it takes to separate the graph into two comparably sized pieces (this can be thought of as the ramification). To calculate the energy barrier in this way, we choose a cut that minimizes the number of crossed bonds in the limit $l\to\infty$, and then compute the total number of bonds that would be frustrated if the spins on one side of the cut were flipped and those on the other not. An example of such a minimal cut for odd $b$ is a vertical line running down the centre of the graph, as shown in \Fref{f:sierpinskigraphcut}. It can directly be computed that the total number of bonds that such a minimal cut crosses is $\left[\frac{(b-c)^{l+1}-1}{(b-c)-1}+1\right]$, providing a lower bound to the energy barrier $\Delta E$. Since the total number of spins in this Ising model is $n\equiv|V(b,c,l)|$, we see that $\Delta E\geq O\left(n^{\frac{1}{\log_{(b-c)}(b^2-c^2)}}\right)$, i.e.~$\Delta E$ is polynomial in $n$ (as the maximum possible energy is also polynomial in $n$, $\Delta E$ cannot be superpolynomial). Though a polynomial energy barrier is typical for models with finite temperature phase transitions such as the 2D Ising model or the 4D toric code, examples exist of systems with polynomial energy barrier but no finite temperature phase transition, such as the welded codes~\cite{Michnicki2012}. It should also be noted that a polynomial energy barrier is necessary for a stabilizer Hamiltonian to have an exponential memory lifetime~\cite{Temme2014, Temme2015}, but not sufficient~\cite{Yoshida2014}.
		
\section{Fractal product codes}\label{s:FPC}

	In this section we define the fractal product codes (FPCs). Although we call them codes, they should be understood as either quantum codes (in the sense of a subspace of a larger Hilbert space) or as local commuting Hamiltonians (such that the frustration-free ground space is the corresponding codespace) depending on context. Since the codes we present are (CSS) stabilizer codes, the Hamiltonian formulation simply corresponds to a negative sum of generators of the stabilizer group. As such, the presentation of the codes will contain more information than is necessary to specify the codespace only, as we will also be interested in the particular (typically non-minimal) choice of generators of the stabilizer group. By setting the Hamiltonian, this choice of generators will set the energetics and thermodynamic properties of the system. Since the code is CSS, we can also consider the $X$ sector and the $Z$ sectors of the code separately as classical codes or classical Hamiltonian systems in the analogous way.
	
	In order to define the FPCs, it will be convenient to recall the homological product of two codes.

	\subsection{Homological product codes}	
	
		The homological (or hypercomplex) product~\cite{Freedman2014, Bravyi2013homological} is a construction for building new CSS codes from existing ones, making use of tools from algebraic topology. We need not introduce the full generality of the homological product here and will simply sketch it as is appropriate for our needs; we refer the interested reader to Refs.~\cite{Freedman2014, Bravyi2013homological}, or a standard reference on algebraic topology, e.g.~\cite{HatcherBook}.
		
		In the homological product construction, each quantum CSS code $\mathcal{C}$ is represented by three vector spaces over the binary field $\F_2$: $\mathcal{C}_0$, $\mathcal{C}_1$, and $\mathcal{C}_2$, and two maps $\partial_{2}^{\mathcal{C}}:\mathcal{C}_2\to\mathcal{C}_1$ and $\partial_{1}^{\mathcal{C}}:\mathcal{C}_1\to\mathcal{C}_0$ such that $\partial_{1}^{\mathcal{C}}\partial_{2}^{\mathcal{C}}=0$. The basis vectors of $\mathcal{C}_1$ correspond to qubits, while those of $\mathcal{C}_0$ ($\mathcal{C}_2$) correspond to $X$-type ($Z$-type) stabilizer generators. Note that these need not be minimal generating sets of stabilizers, and in fact the choice of generators significantly affects the construction. The maps $\partial_{2}^{\mathcal{C}}$ and $\left(\partial_{1}^{\mathcal{C}}\right)^T$ define the qubits on which each stabilizer generator has support, and the constraint $\partial_{1}^{\mathcal{C}}\partial_{2}^{\mathcal{C}}=0$ enforces that the stabilizer group is abelian.
		
		Often, as will be the case in our construction, the basis vectors of $\mathcal{C}_i$ correspond to geometrical objects of dimension $i$, e.g.~vertices, edges, and plaquettes of some cell complex for $i=0,1,2$ respectively. The maps $\partial_i^{\mathcal{C}}$ then specify the $(i-1)$-dimensional objects that comprise the boundary of a given $i$-dimensional object. This description can be understood as a generalized toric code construction, where qubits are placed on $i$-dimensional objects, $X$-type stabilizers associated with $(i-1)$-dimensional objects, and $Z$-type stabilizers associated with $(i+1)$-dimensional objects.
		
		Associated with each space $\mathcal{C}_i$ is a homology group $H_i(\mathcal{C})=\ker\partial^{\mathcal{C}}_i/\im\partial^{\mathcal{C}}_{i+1}$ and cohomology group $H^i(\mathcal{C})=\ker\left(\partial^{\mathcal{C}}_i\right)^T/\im\left(\partial^{\mathcal{C}}_{i-1}\right)^T$ (with $\partial_3^\mathcal{C}$ and $\partial_0^{\mathcal{C}}$ maps from and to the zero space, respectively, such that $\im\partial^{\mathcal{C}}_3=\im\left(\partial^{\mathcal{C}}_0\right)^T=0$, $\ker\partial^{\mathcal{C}}_0=\mathcal{C}_0$, and $\ker\left(\partial^{\mathcal{C}}_3\right)^T=\mathcal{C}_3$). In this language, the $X$-type and $Z$-type logical operators correspond to elements of $H_1(\mathcal{C})$ and $H^1(\mathcal{C})$ respectively. As such, the number of encoded qubits in such a code is given by $k_{\mathcal{C}}\equiv \dim H_1(\mathcal{C})=\dim(\ker\partial^{\mathcal{C}}_1)-\dim(\im\partial^{\mathcal{C}}_2)$ (or equivalently $\dim H^1(\mathcal{C})$).
		
		The homological product of two codes $\mathcal{C}$ and $\mathcal{C'}$ yields a new object $\mathcal{C}\otimes\mathcal{C'}$ with five spaces and four maps, given by
		\begin{align}
			\left(\mathcal{C}\otimes\mathcal{C'}\right)_j &= \bigoplus_{i=0}^j \left(\mathcal{C}_i\otimes\mathcal{C'}_{i-j}\right)\\
			\partial^{\left(\mathcal{C}\otimes\mathcal{C'}\right)}_{j}\left(c_i\otimes c_{j-i}'\right) &= \left(\partial^{\mathcal{C}}_ic_i\right)\otimes c'_{j-i} + c_{i}\otimes \left(\partial^{\mathcal{C'}}_ic_{j-i}'\right)
		\end{align}
		for $c_i \in \mathcal{C}_i$ and $c'_i\in \mathcal{C'}_i$.
		
		We could define three different codes from the general construction, but for our purposes we take the middle homological product code, denoted by $\mathrm{mid}\left(\mathcal{C}\otimes \mathcal{C'}\right)$ and defined by the three spaces $\left(\mathcal{C}\otimes\mathcal{C'}\right)_1$, $\left(\mathcal{C}\otimes\mathcal{C'}\right)_2$, $\left(\mathcal{C}\otimes\mathcal{C'}\right)_3$, and two maps $\partial^{\left(\mathcal{C}\otimes\mathcal{C'}\right)}_{2}$, and $\partial^{\left(\mathcal{C}\otimes\mathcal{C'}\right)}_{3}$. In $\mathrm{mid}\left(\mathcal{C}\otimes\mathcal{C'}\right)$, basis vectors of the space $\left(\mathcal{C}\otimes\mathcal{C'}\right)_2$ correspond to qubits, while basis vectors of the spaces $\left(\mathcal{C}\otimes\mathcal{C'}\right)_1$ and $\left(\mathcal{C}\otimes\mathcal{C'}\right)_3$ correspond to $X$- or $Z$-type stabilizer generators, respectively.
		
		Properties of the homological product codes can be determined directly from those of their component codes. In particular, it will be useful to determine the logical operators of such codes. These can be calculated using the K\"unneth formulae
		\begin{align}
			H_i\left(\mathcal{C}\otimes\mathcal{C'}\right)&=\bigoplus_{\substack{j,k:\\j+k=i}}H_j(\mathcal{C})\otimes H_k(\mathcal{C'})\\
			H^i\left(\mathcal{C}\otimes\mathcal{C'}\right)&=\bigoplus_{\substack{j,k:\\j+k=i}}H^j(\mathcal{C})\otimes H^k(\mathcal{C'})
		\end{align}
		The logical operators of the middle homological product code correspond to the elements of the homology and cohomology groups $H_2\left(\mathcal{C}\otimes\mathcal{C'}\right)$ and $H^2\left(\mathcal{C}\otimes\mathcal{C'}\right)$, and correspondingly the number of encoded qubits is given by $k_{\mathrm{mid}\left(\mathcal{C}\otimes\mathcal{C'}\right)}=\dim H_2\left(\mathcal{C}\otimes\mathcal{C'}\right)$.

	\subsection{Defining the FPCs}\label{s:fpcdef}
		
	Consider the toric code~\cite{Kitaev2003} (or more properly the surface code~\cite{Dennis2002}) $\mathcal{T}(b,c,l)$ on a Sierpinski carpet graph $\widehat{SC}(b,c,l)$ defined in the standard way, so that qubits reside on the edges and $X$- or $Z$-type stabilizers are associated with vertices or plaquettes of the graph, respectively. These are conventionally given as $A_v=\prod_{e\sim v}X_e$ and $B_p=\prod_{e\sim p}Z_e$ for $X_e$ and $Z_e$ the Pauli operators on edge $e$. We only associate $Z$-type stabilizers with the interior plaquettes, so as to retain locality of the stabilizer generators.
	
	This is a surface code with holes punched into the manifold for each independent exterior plaquette, each of these holes and the outer boundary having smooth boundary conditions. Since such a system with $h$ smooth holes plus the outer smooth boundary encodes $h$ qubits~\cite{Dennis2002}, and the Sierpinski carpet graph $\widehat{SC}(b,c,l)$ has $|P_e(b,c,l)|$ exterior plaquettes not including the outer boundary, it is clear that the (logarithm of the) degeneracy of this code is $k_{\mathcal{T}}=|P_e|=\frac{1-(b^2-c^2)^l}{1-(b^2-c^2)}$ (we suppress the $(b,c,l)$ labels as convenient). It can also be directly verified that the number of qubits in the system is equal to the total number of independent stabilizers, plus the number of encoded qubits, i.e.~$|E|=|P_i|+|V|-1+|P_e|$, where we have noted that there are $|V|-1$ independent $X$-type stabilizers.
		
		In the homological language, the spaces $\mathcal{T}_0$, $\mathcal{T}_1$, and $\mathcal{T}_2$ correspond to vertices, edges, and interior plaquettes of $\widehat{SC}(b,c,l)$ respectively. The homology group $H_0(\mathcal{T})$ is given as the quotient of the space of vertices by the space of vertices that are boundaries of some set of edges. Since all sets containing an even number of vertices are boundaries, we see that $H_0(\mathcal{T})\cong \Z_2$ breaks into even and odd elements. In contrast, the second homology group $H_2(\mathcal{T})$ is given as the sets of plaquettes with no boundary, which is empty, giving $H_2(\mathcal{T})\cong 0$. The first homology group is given by the quotient of the cycles in the graph by the interior boundaries. Since there are $|P_e|=k_{\mathcal{T}}$ independent exterior boundaries, we find $H_1(\mathcal{T})\cong \Z_2^{k_{\mathcal{T}}}$ as expected.
		
\begin{figure}
			\begin{subfigure}{0.23\textwidth}
			\includegraphics{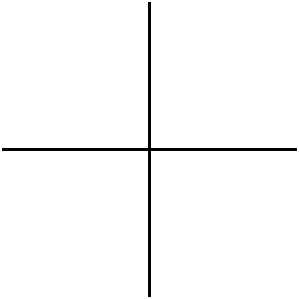}
			\end{subfigure}
			\begin{subfigure}{0.23\textwidth}
			\includegraphics{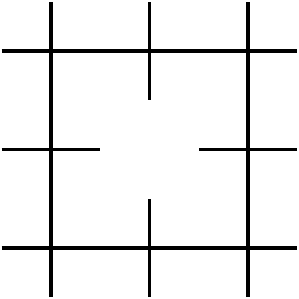}
			\end{subfigure}
			\begin{subfigure}{0.23\textwidth}
			\includegraphics{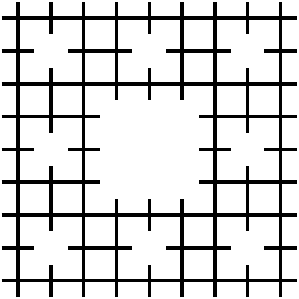}
			\end{subfigure}
			\begin{subfigure}{0.23\textwidth}
			\includegraphics{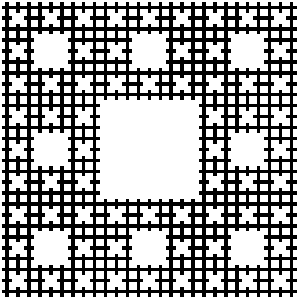}
			\end{subfigure}
			\caption{$\widehat{SC}^*(3,1,l)$ for $l=0$ to $3$, neglecting the exterior vertices.}\label{f:sierpinskigraphdual}
		\end{figure}

		We will also make use of the dual code $\mathcal{T}^*$, where the $X$-type and $Z$-type stabilizers have been exchanged. This is the toric code on the dual graph to $\widehat{SC}(b,c,l)$ (i.e.~where plaquettes and vertices have been exchanged, and which we denote $\widehat{SC}^*(b,c,l)$, see \Fref{f:sierpinskigraphdual}), with the appropriate high-coordination vertices neglected for the purposes of defining the stabilizer group. In the homological language, a dual code has $\mathcal{C}^*_0=\mathcal{C}_2$, $\mathcal{C}^*_1=\mathcal{C}_1$, $\partial^{\mathcal{C^*}}_1=\left(\partial^{\mathcal{C}}_2\right)^T$ and $\mathcal{C}^{**}=\mathcal{C}$. From these duality properties, the homology groups of this code are immediate: $H_2(\mathcal{T^*})\cong \Z_2$, $H_0(\mathcal{T^*})\cong 0$, and $H_1(\mathcal{T})\cong \Z_2^{k_{\mathcal{T}}}$.

		We define the family of fractal product codes $FPC(b,c,l)$ as the middle homological product $\mathrm{mid}\left(\mathcal{T}\otimes \mathcal{T}^*\right)$ for suitable $b$ and $c$. The physical qubits of this code correspond to basis vectors of $\left(\mathcal{T}\otimes\mathcal{T}^*\right)_2=\mathcal{T}_0\otimes\mathcal{T}^*_2\oplus\mathcal{T}_1\otimes\mathcal{T}^*_1\oplus\mathcal{T}_2\otimes\mathcal{T}^*_0$. There are thus $n_{FPC(b,c,l)}\equiv |V(b,c,l)|^2+|E(b,c,l)|^2+|P_i(b,c,l)|^2$ physical qubits in this code.
		
			The degeneracy of these codes can easily be determined by the K\"unneth formula as
			\begin{align}
				k_{FPC(b,c,l)}&=\dim \left(H_0(\mathcal{T})\otimes H_2(\mathcal{T}^*)\right)+\dim \left(H_1(\mathcal{T})\otimes H_1(\mathcal{T}^*)\right)+\dim \left(H_2(\mathcal{T})\otimes H_0(\mathcal{T}^*)\right)\\
				&=\left(1\cdot 1\right)+\left(k_{\mathcal{T}}\right)^2+\left(0\cdot 0\right)\\
				&=1+\left(\frac{1-(b^2-c^2)^l}{1-(b^2-c^2)}\right)^2\label{e:kunneth}
			\end{align}	
			where, in the second last line, we note that $H_2(\mathcal{T}^*)\cong H_0(\mathcal{T})\cong\Z_2$ and $H_2(\mathcal{T})\cong H_0(\mathcal{T}^*)\cong 0$.
			
			The number of qubits in the code $n_{FPC(b,c,l)}$ scales asymptotically as $(b^2-c^2)^{2l}$, as does the number of encoded qubits $k_{FPC(b,c,l)}$, suggesting a constant rate $r_{FPC(b,c,l)}\equiv\frac{k_{FPC(b,c,l)}}{n_{FPC(b,c,l)}}$.	

			The degeneracy of this system comes from two different sources. The $H_1(\mathcal{T})\otimes H_1(\mathcal{T}^*)$ term gives rise to an extensive number of encoded qubits $\left(\frac{1-(b^2-c^2)^l}{1-(b^2-c^2)}\right)^2$, and we will call these qubits ``local''. By contrast, the $H_0(\mathcal{T})\otimes H_2(\mathcal{T}^*)$ term produces a single qubit of degeneracy, and we call this qubit ``global''. The two kinds of encoded qubits have quite different properties, and we will largely focus on the global encoded qubit.
			
			Some intuition for the properties of an FPC can be gained by considering its relation to the 4D toric code. Noting that both $\widehat{SC}(b,c,l)$ and $\widehat{SC}^*(b,c,l)$ are subgraphs of the square lattice, we see that $\widehat{SC}(b,c,l)\times\widehat{SC}^*(b,c,l)$ is a subgraph of the 4D hypercubic lattice. The qubits of the FPC correspond to basis vectors of $\left(\mathcal{T}\otimes\mathcal{T}^*\right)_2$, i.e.~(interior) plaquettes of $\widehat{SC}(b,c,l)\times\widehat{SC}^*(b,c,l)$. Similarly the $X$-type and $Z$-type stabilizers correspond to the links $\left(\mathcal{T}\otimes\mathcal{T}^*\right)_1$ and cubes $\left(\mathcal{T}\otimes\mathcal{T}^*\right)_3$ respectively. When interpreting an FPC in this way, it appears as the 4D toric code~\cite{Dennis2002} defined on a fractal subset of $\Z^4$. In order to complete the specification we must determine the boundary conditions at each boundary of this surface. If we consider the graph $\widehat{SC}(b,c,l)$ to be oriented in the $\hat{x}\text{--}\hat{y}$ plane, and extended by $\widehat{SC}^*(b,c,l)$ into the $\hat{w}\text{--}\hat{z}$ plane, we should consider boundaries with normals in the $\hat{x}$ or $\hat{y}$ directions to be ``smooth'' and those in the $\hat{w}$ or $\hat{z}$ directions to be ``rough''. It is straightforward to show that the total number of qubits in this code scales asymptotically as $L^{2\dimh SC(b,c)}$ for $L$ the linear lattice size of the hypercubic lattice.
			
			As previously mentioned, the encoded qubits of this code can be classified as either global or local. Interpreting the system as a punctured 4D toric code, we can see why this characterization has been made. As can be seen by studying the homology and cohomology groups $H_2\left(\mathcal{T}\otimes\mathcal{T}^*\right)$ and $H^2\left(\mathcal{T}\otimes\mathcal{T}^*\right)$, each local qubit has corresponding logical operators whose support can be localized on a membrane around and between punctures in the 4D toric code, while the global qubit has corresponding logical operators that run in either the $\hat{x}\text{--}\hat{y}$ plane or the $\hat{w}\text{--}\hat{z}$ plane, and so have support on a membrane in the shape of the Sierpinski carpet $SC(b,c,l)$. These global logical operators inherit much of the structure of the Ising model on the Sierpinski carpet, and in particular their polynomial energy barrier.
		
	It is clear that the FPCs are self-dual in the sense that by exchanging the $\hat{x}$ and $\hat{y}$ axes with the $\hat{w}$ and $\hat{z}$ axes we exchange the $X$ and $Z$ sectors of the code. Algebraically, this can be seen from the fact that the homological products $\mathcal{C}\otimes\mathcal{C}'$ and $\mathcal{C}'\otimes\mathcal{C}$ are isomorphic.
		
		
\section{Thermodynamics of FPCs}

	The Hamiltonian for an FPC is given by a negative sum of stabilizer generators for each link and cube of the lattice:
	\begin{align}
		H_{FPC}(b,c,l)&= -\sum_{A\in\left(\mathcal{T}\otimes\mathcal{T}^*\right)_1}\prod_{j\in\left(\partial^{\left(\mathcal{T}\otimes\mathcal{T}^*\right)}_2\right)^TA}X_j
		 -\sum_{B\in\left(\mathcal{T}\otimes\mathcal{T}^*\right)_3}\prod_{k\in\partial^{\left(\mathcal{T}\otimes\mathcal{T}^*\right)}_3B}Z_k
	\end{align}
	with $X_i$ and $Z_i$ the relevant Pauli matrices on qubit $i$.	
	
	In order to study the thermodynamic properties of the quantum Hamiltonians corresponding to $FPC(b,c,l)$, it will be convenient to consider each of the two sectors $X$ and $Z$ individually. Since the $X$-type and $Z$-type stabilizers commute pairwise, for the purposes of considering thermalization processes we can consider each sector separately (see Ref.~\cite{Alicki2010} for an analogous discussion). We denote the corresponding classical codes $FPC_X(b,c,l)$ and $FPC_Z(b,c,l)$ with associated Hamiltonians
	\begin{align}
		H_{FPC_X}(b,c,l)&= -\sum_{A\in\left(\mathcal{T}\otimes\mathcal{T}^*\right)_1}\prod_{j\in\left(\partial^{\left(\mathcal{T}\otimes\mathcal{T}^*\right)}_2\right)^TA}X_j\\
		H_{FPC_Z}(b,c,l)&= -\sum_{B\in\left(\mathcal{T}\otimes\mathcal{T}^*\right)_3}\prod_{k\in\partial^{\left(\mathcal{T}\otimes\mathcal{T}^*\right)}_3B}Z_k
	\end{align}	
	As noted, these two systems are related by a rotation and so can be considered dual to one another, meaning we need only study the properties of one of these classical codes. The thermodynamic limit corresponds to taking $l\to\infty$ for fixed $b$ and $c$.
	
	The main technical result of this paper is the following:

	\begin{theorem}\label{t:phasetransition}
		The classical Hamiltonians corresponding to each of the classical codes $FPC_X(b,c)$ and $FPC_Z(b,c)$ have finite temperature phase transitions.
	\end{theorem}

	The proof of this theorem, given in \Sref{s:thmproof}, makes use of two main tools: duality transformations and correlation inequalities. Both of these results apply to the class of generalized Ising models, to which $FPC_X$ and $FPC_Z$ belong. Generalized Ising models consist of ferromagnetic interactions on sets of spins (for simplicity, spin-$\frac{1}{2}$). Such systems are specified by a lattice of spins $\Lambda$, and interaction strengths $J_{R}\geq 0$ for each $R\subset \Lambda$, with a Hamiltonian
	\begin{align}
		H(\{J_R\}_R)=-\sum_{R\subseteq \Lambda} J_{R}\prod_{j\in R}Z_j
	\end{align}
	It is clear that each of the sectors of the FPC can be trivially written in this way (up to a trivial change of basis).
	
	We will be interested in the phase transitions of such a model at finite inverse temperature $\beta$, and so it will be convenient to define rescaled interaction strengths $K_R=\beta J_R$, the set of non-trivial interactions $\mathcal{K}=\{K_R>0\}$, and their supports $\mathcal{B}=\{R\subseteq \Lambda|K_R\in\mathcal{K}\}$. We will also abuse notation and treat suitable sets as groups when convenient, with multiplication given as the exclusive union $g_1g_2=\left(g_1\cup g_2\right) \setminus \left(g_1\cap g_2\right)$ for $g_i$ in an appropriate set such as $\mathcal{B}$. Also of interest is the symmetry set (or group) $\mathcal{S}$, with elements $\mathcal{S}=\{R|(-1)^{|S\cap B|}=1 \;\forall B\in\mathcal{B}\}$, and $N_\mathcal{S}$, defined as the number of generators of the group $\mathcal{S}$, so that $|\mathcal{S}|=2^{N_\mathcal{S}}$. $\mathcal{S}$ should be understood physically as the group of spin flips that commute with each term in the Hamiltonian. Finally, it is convenient to construct the group of constraints $\mathcal{C}=\{C\subseteq \mathcal{B}|\prod_{c_i\in C}\prod_{j\in c_i}Z_j=I\}$. Elements of this group are sets of interactions that are not independent.
	
	The partition function of a generalized Ising model is
	\begin{align}
		\mathcal{Z}(\Lambda, \mathcal{K})=\sum_{R\subseteq \Lambda}\exp\left(\sum_{B\in\mathcal{B}}K_B(-1)^{|R\cap B|}\right)
	\end{align}
	where the $R$ represents possible sets of spins pointing down, and so $\prod_{j\in B}Z_j$ acting on the corresponding state will accumulate a phase $(-1)^{|R\cap B|}$. 
	
	\subsection{Correlation inequalities} \label{s:gks}
	
		The GKS inequalities~\cite{Griffiths1967, Kelly1968} (see also Ref.~\cite{GriffithsBook}) are simple correlation inequalities for ferromagnetic systems. These inequalities state that for a generalized Ising model,
		\begin{align}
			\sizedexpect{\left(\prod_{j\in A\subset \Lambda}Z_j\right)\left(\prod_{j'\in B\subset \Lambda}Z_{j'}\right)}\geq \sizedexpect{\left(\prod_{j\in A}Z_j\right)}\sizedexpect{\left(\prod_{j'\in B}Z_{j'}\right)}
		\end{align}
		An immediate corollary of this inequality is that
		\begin{align}
			\frac{\partial}{\partial K_B}\sizedexpect{\left(\prod_{j\in A\subset \Lambda}Z_j\right)}&= \sizedexpect{\left(\prod_{j\in A}Z_j\right)\left(\prod_{j'\in B}Z_{j'}\right)}- \sizedexpect{\left(\prod_{j\in A}Z_j\right)}\sizedexpect{\left(\prod_{j'\in B}Z_{j'}\right)}\geq 0
		\end{align}
		Intuitively, increasing the strength of, or adding more ferromagnetic interactions cannot decrease the correlations present in the system. In particular, if there exists an ordered phase of a ferromagnetic system, it cannot be destroyed by adding extra symmetry-respecting ferromagnetic terms to the Hamiltonian. This is the crucial sense in which we will make use of the GKS inequalities.
		
	\subsection{Duality transformations on general Ising models}
	
		Duality transformations are extensively used in statistical mechanics to relate the thermodynamic properties of different models. They have previously been used to study the properties of topologically ordered models~\cite{Nussinov2008,Nussinov2012}. We make use of a particular family of duality transformations on general Ising models due to Merlini and Gruber~\cite{Merlini1972} that have a natural geometrical interpretation. Given a system of spins $\Lambda$ and set of interactions $\mathcal{K}$, Merlini and Gruber give a prescription to construct a dual system $\Lambda^*$ with interactions $\mathcal{K}^*$ and a (surjective) map $\varphi: \mathcal{B}\to\mathcal{B}^*$ such that
		\begin{align}
			\mathcal{Z}(\Lambda,\mathcal{K})&=\sqrt{2}^{(|\Lambda|-|\Lambda^*|+N_\mathcal{S}-N_\mathcal{S}^*)}\prod_{B\in\mathcal{B}}\sqrt{\mathrm{\sinh}2K_B}\cdot \mathcal{Z}(\Lambda^*,\mathcal{K}^*)
		\end{align}
		for $\e^{-2K^*_{B^*}}=\prod_{B\in \varphi^{-1}B^*}\tanh K_B$. This relation between the partition functions ensures that a non-analyticity in the free energy (i.e.~a phase transition) in the $(\Lambda^*,\mathcal{K}^*)$ system also corresponds to a phase transition in the $(\Lambda,\mathcal{K})$ system at an appropriately rescaled temperature.
		
		In order to construct such a dual system, we consider an arbitrary generating set of the constraint group $\mathcal{C}$. For each generator $C_i$, we assign a site $v_i^*\in\Lambda^*$. The interaction regions $B^*_j$ are labelled by interaction terms $B_j\in\mathcal{B}$, with $B^*_j=\{v_i^*\in\Lambda^*|B_j\in C_i\}$.
		
	\subsection{Finite temperature phase transition}\label{s:thmproof}
	
		Given the Merlini-Gruber duality transformation and the GKS inequalities, we can now prove \Tref{t:phasetransition}.
		
		Consider $FPC_Z(b,c,l)$. The set of spins in this model $\Lambda$ are associated with (a subset of the) faces of the hypercubic lattice. Of course, this is according to their presence in $\left(\mathcal{T}\otimes\mathcal{T}^*\right)_2$, or equivalently $G\equiv\widehat{SC}(b,c,l)\times\widehat{SC}^*(b,c,l)$ modulo exterior boundaries (we will also use $G$ to label this corresponding lattice, where it is understood that exterior boundaries as neglected as appropriate). The interaction regions $\mathcal{B}$ are associated with cubes of $G$, and each includes the faces that bound it. We perform a Merlini-Gruber duality transformation on this system as follows:
		
		The constraint group of $FPC_Z$ can be generated by elements labelled by hypercubes (formed by all cubes bounding the hypercube) and exterior boundaries (formed by all smooth cubes terminating at that boundary) of the lattice $G$. Thus to determine the dual system, we place sites at each hypercube and exterior boundary of $G$, and we determine the interactions of the dual system by noting that each cube interaction in the bulk belongs to two hypercubes, and at the (smooth) boundaries each cube interaction belongs to a single hypercube constraint and the relevant exterior boundary constraint. Thus the interactions of the dual system will all be 2-body. Though it is relatively clear what the geometry of such a system looks like, we need not consider the entire lattice. Instead, it suffices to consider a single slice in the $\hat{w}\text{--}\hat{z}$ plane. 
		
		If we consider extending the graph $\widehat{SC}^*(b,c,l)$ (lying in the $\hat{w}\text{--}\hat{z}$ plane) by a single plaquette from $\widehat{SC}(b,c,l)$, we find hypercubes corresponding to each plaquette of $\widehat{SC}^*(b,c,l)$. Since the dual system replaces hypercubes by vertices, and has nearest-neighbour 2-body interactions, the corresponding slice of the dual system gives an Ising model on the planar dual graph to $\widehat{SC}^*(b,c,l)$, that is, $\widehat{SC}(b,c,l)$. Thus, the Hamiltonian of the dual system will consist of an Ising model on $\widehat{SC}(b,c,l)$ plus some other terms, i.e.
		\begin{align}
			H(\Lambda^*,\mathcal{K}^*)=H_{SC}+\ldots\label{e:dualsplit}
		\end{align}
		for $H_{SC}$ the ferromagnetic Ising model on the relevant $(b,c,l)$ Sierpinski carpet graph as described in \Sref{s:scising}. Since the symmetry group of the dual model is simply generated by $\prod_{v^*\in\Lambda^*}X_{v^*}$, and all terms neglected in (\ref{e:dualsplit}) are 2-body, it is clear that they respect this symmetry.
		
		Now appealing to the GKS inequality, if we were to begin with the system $H_{SC}$, then by adding the additional symmetry-respecting ferromagnetic terms to give $H(\Lambda^*,\mathcal{K}^*)$ we cannot decrease the (spontaneous) magnetization of the system. Since it is known that the Ising model on a Sierpinski carpet $H_{SC}$ has a magnetically ordered phase for some range of finite inverse temperatures $\beta$~\cite{Shinoda2002, Vezzani2003, Campari2010}, this tells us that $H(\Lambda^*,\mathcal{K}^*)$ will also have such an ordered phase. Although there are infinitely many dual interactions acting on each of the spins corresponding to exterior boundaries of $G$, in the bulk $H(\Lambda^*,\mathcal{K}^*)$ acts with a bounded density of bounded strength operators. We therefore conclude that in the $\beta\to 0$ limit the system will be disordered, demonstrating that there must be a finite temperature phase transition.
		
		Since $H(\Lambda^*,\mathcal{K}^*)$ is dual to the Hamiltonian $H_{FPC_Z}$, we deduce that it too possesses a finite temperature phase transition. The symmetry between $FPC_Z$ and $FPC_X$ implies that the same is true for the $X$ sector of the theory, completing the proof of \Tref{t:phasetransition}.
		
		Though the following is by no means a rigorous argument, we intuitively associate the identified phase transitions with the global encoded qubit, as the Hamiltonian $H_{SC}$ in Eq.~(\ref{e:dualsplit}) corresponds only to action on those qubits of $H_{FPC_Z}$ in a cross-sectional slice of the lattice, as do the logical operators of the global encoded qubit. The existence of finite temperature phase transitions such as those identified in \Tref{t:phasetransition} is indicative that the system may be able to robustly store quantum information in this global qubit at finite temperature.


\section{Revisiting the Caltech Rules}

	By construction, FPCs satisfy the non-trivial codespace Caltech rule, as has been shown using the K\"unneth formula in \Eref{e:kunneth}. We will now discuss the remaining rules in turn.
	
	\subsection{Embedding into $\R^3$ (finite spins, bounded local interactions)}
	
		The fractal lattice on which an FPC is defined is a discretization of $SC(b,c)\times SC(b,c)$. By choosing the parameters $b$ and $c$ such that $\dimh SC(b,c)\leq \frac{3}{2}$, this fractal will have Hausdorff dimension less than or equal to 3. Though this may be enough by itself to warrant curiosity about the properties of such a model, it is clearly a desirable feature that these codes can be implemented with local interactions in $\R^3$.
		
		Most studies of embeddings of fractals into other spaces are concerned with bilipschitz embeddings. Such an embedding would neither increase distances (making previously local interactions nonlocal) or decrease distances (increasing the density of sites) by more than a constant factor. It seems unlikely that a bilipschitz embedding is possible from $SC(b,c)\times SC(b,c)$ to $\R^3$, no matter what values of $b$ and $c$ are chosen. However, such an embedding is not necessary to preserve locality. In particular, it is possible for a map to take a finite number of far separated points to the same location, and still preserve locality in the sense that we require. This is because such an map will only increase the density of points by a constant factor. As an example, a projection of the torus into the plane will map two distant points to the same location, but in doing so will only increase the density of sites by a factor of 2. This would still allow the model to satisfy the relevant Caltech rules. Furthermore, the fact that the sites of our code only lie on a discrete lattice, instead of in over the entire fractal as would normally be considered, may provide further flexibility in constructing a locality-preserving map into $\R^3$.
		
		While we leave the realization of FPCs in $\R^3$ as an open question, we will briefly discuss some relevant facts that are suggestive of this possibility. It is well known that a random projection of a fractal with dimension $d<D$ into $\R^D$ will yield an object with dimension $d$~\cite{FalconerBook}. By itself, this is suggestive that certain features of a suitable fractal survive under a random projection. Clearly a projection will not increase the distance between qubits, however one must also be wary that this projection will not cause the density to diverge.
		
		In order to consider whether or not this will happen for a random projection of $SC(b,c)\times SC(b,c)$ into $\R^3$, it is interesting to consider a related family of fractals with low \emph{lacunarity}. Lacunarity is a measure of violation of translation invariance, and low-lacunarity Sierpinski carpets can be constructed by varying the location of the removed volumes at each level~\cite{Gefen1983}. At a fixed Hausdorff dimension, the lacunarity can be made arbitrarily small, in which case the density of points in the fractal approaches uniformity. If the Hausdorff dimension is less than 3, then a random projection of a fractal with vanishing lacunarity into $\R^3$ should lead to a bounded density as required. Using these ideas, one could either directly define a family of low-lacunarity FPCs that can be locally realized in $\R^3$, or it may be possible to extrapolate from this result to demonstrate a local realization in $\R^3$ of the family of FPCs defined in \Sref{s:fpcdef}.
		
		Complicating this simple idea is the fact that that taking both the low lacunarity limit and the thermodynamic limit of our discrete lattice together at a fixed Hausdorff dimension seems quite a subtle task. Another obstacle is that constructions of low lacunarity fractals are not unique or canonical. In addition, although it is expected that Ising models on low-lacunarity Sierpinski carpets will have finite temperature phase transitions~\cite{Gefen1983}, the effect of lacunarity on the thermodynamic properties such as the critical temperature of Ising models is not well understood~\cite{Gefen1984III, Wu1987}. Another related class of fractals that may be instructive for considering projections into $\R^3$ are the random Sierpinski carpets~\cite{Perreau1996}, although their properties are also poorly understood.

	\subsection{Perturbative stability} \label{s:pertstab}
	
		Typically, perturbative stability of the codespace of a system similar to an FPC is shown by proving that the gap and ground space degeneracy is stable under arbitrary quasi-local perturbations, as in the topological stability theorems~\cite{Bravyi2010, Bravyi2011, Michalakis2013}. This guarantees that a quasi-adiabatic continuation between the ground spaces of the perturbed and unperturbed models exists~\cite{Bravyi2010, Bachmann2012}, ensuring that their properties are stable. Though $H_{FPC}$ are gapped, the topological stability theorems do not apply to our system, as the ground space does not have macroscopic distance (due to the presence of the local encoded qubits). Under generic perturbation, naive perturbation theory suggests that the degeneracies associated with the local encoded qubits will be lifted, while splitting of levels associated with the global encoded qubit will be exponentially suppressed in system size as desired (due to the local indistinguishability of the global qubit states).
	
	In order to make the stability of the global encoded qubit more concrete, a finer notion of perturbative stability seems to be required. We do not expect that our entire ground space will be stable, nor a subspace of it. Instead, it seems likely that by factorizing the ground space into the global encoded qubit system and the local encoded qubit systems, the global subsystem will be stable in the sense that an appropriate continuation could be constructed from the unperturbed logical space to the perturbed logical space. Similar considerations would apply to other systems that might be of independent interest, such as a toric code with a small but non-zero density of punctures, or a nonabelian anyon model with a finite density of particles.
	
	Since the specification of which perturbations are local is dependent on the embedding of the code into Euclidean space, one might wonder whether the perturbative stability properties of an FPC could be different depending on whether it is realized in $\R^4$, or in $\R^3$ (assuming that this is possible). However, since the relevant arguments in the topological stability theorems do not directly depend on the embedding, only on the properties of the code itself such as the distance scaling polynomially with $L$, we do not expect that this will be the case.

	\subsection{Efficient decoding}
	
		Due to the similarity between the 4D toric code and the FPCs, it seems likely that any decoding algorithm for the 4D toric code may be adapted to decode the FPCs. A notable example of such a decoder that also applies to topological codes in general is the topological renormalization group decoder due to Bravyi and Haah~\cite{Bravyi2011b,Bravyi2013}. Unfortunately, for the same reasons that we could not apply the topological stability theorems (namely the presence of local encoded qubits), the proof of threshold for this decoder does not directly apply for the FPCs. However, we anticipate that this algorithm and proof of threshold could be adapted to our setting. Given the self-similar nature of the FPCs, alternative renormalization group based decoding methods~\cite{Duclos-Cianci2010,Duclos-Cianci2010a,Duclos-Cianci2013} might also be effective decoders for these systems. Though we would not expect identical thresholds, it may also be possible to derive optimal thresholds for FPCs using similar statistical mechanical tools as have been applied to the 4D toric code~\cite{Takeda2004}.
		
		For the purposes of decoding the global encoded qubit, the FPC should be treated as a subsystem code~\cite{Kribs2005, Kribs2006}, with the local encoded degrees of freedom playing the role of gauge qubits. Other natural candidates for an FPC decoding algorithm are the 4D toric code heat-bath algorithm~\cite{Dennis2002} or a variant of Toom's rule~\cite{Toom1980}. Particularly, a concrete proof that the FPC is indeed a self-correcting memory would likely also guarantee the existence of a polynomial decoding algorithm, as in Ref.~\cite{Alicki2010} for the 4D toric code. For similar reasons, we also expect that the FPCs may be single-shot fault-tolerant~\cite{Bombin2014single}.

	\subsection{Exponential lifetime}
		
		We have not proven that the memory lifetime of a FPC scales exponentially with system size for some finite temperatures, simply that there are phase transitions of the system at finite temperature. This seems highly suggestive that the system may function as a self-correcting quantum memory below the critical temperatures, but the relation between thermodynamic phase transitions and memory lifetime is not fully understood~\cite{Hastings2014,Yoshida2014} and so a rigorous proof of this fact is desirable. We anticipate that it may be possible to construct such a proof by combining the techniques used to prove the exponential lifetime of the 4D toric code~\cite{Alicki2010} (see also Ref.~\cite{Bombin20136d}) with those used in the Peierls-type proof of the finite-temperature phase transition in the Sierpinski carpet Ising model~\cite{Vezzani2003,Campari2010}.

\section{Discussion}

	\subsection{Alternative fractal product codes}
	
		The main ideas of this construction are clearly not limited to the Sierpinski carpet fractals. The results and methods apply directly to any pair of well-behaved fractals (e.g.~self-similar Borel sets satisfying the open set condition) with infinite ramification, corresponding Ising models with finite temperature phase transitions, and combined dimension $\leq 3$. For these fractals an FPC can be constructed in an analogous way, and may also act as a self-correcting quantum memory in 3D. Similarly, by varying the relative dimension of the two fractals used, one can introduce an asymmetry between the critical temperatures of the $X$ and $Z$ sectors of the system. General results are known on the existence of phase transitions in fractal graphs~\cite{Gefen1980, Gefen1984III, Campari2010}, and families of infinitely ramified fractals are known~\cite{Bandt2010}, though these are not as well studied as the Sierpinski carpets treated here.
		
		It should also be noted that the particular discretization of the fractal used to define the fractal graph does not play a critical role in our construction. An alternative convention for defining the Sierpinski carpet graphs, for example as discussed in Refs.~\cite{Monceau1998,Bonnier1987}, would be equally amenable to our analysis.
				
		Though our construction seems to manifestly break translation invariance, it might also be possible to restore it approximately. There is some evidence that variants of the Sierpinski carpet with low lacunarity behave like concrete geometric realizations of hypercubic lattices with fractional dimension in certain limits~\cite{Gefen1983}. It would be interesting to see whether the use of these fractal graphs would allow for a translation-invariant FPC as a limiting case. In order to answer such questions, the precise details of any embedding into $\R^3$ would need to be investigated.
		
	\subsection{Non-fractal product codes}\label{s:nonfractal}			
		
		In addition to the possibility of using alternative fractal graphs to define other families of FPCs, many of our techniques would apply to suitable non-fractal (non-self-similar) graphs. The two main features of the Sierpinski carpet graphs that have been important are that an Ising model on the graph has a finite temperature phase transition, and that the total number of sites of the graph is growing slower than $L^{\frac{3}{2}}$ for $L$ the linear lattice size. We can also consider arbitrary finitely coordinated graphs with finite temperature Ising model phase transitions and suitably slowly growing number of sites.
		
\begin{figure}
			\includegraphics{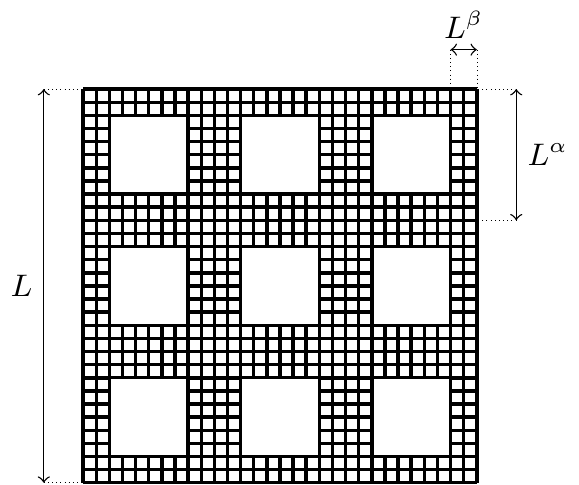}
			\caption{A non-fractal graph that could be used as the basis for a product code can be formed by removing regions from an $L\times L$ square lattice. $\alpha$ and $\beta$ can then be chosen such that the total number of sites of this graph grows slower than $L^{\frac{3}{2}}$, without requiring the self-similarity of a fractal graph.}\label{f:nonfractal}
		\end{figure}

		A family of graphs with these properties can be constructed, for example, by taking an $L\times L$ square lattice and dividing it into squares of size $L^{\alpha}\times L^{\alpha}$ for some $0<\alpha <2$. Central regions from within each of these squares would then be removed to leave only a border of width $L^{\beta}$ for some $0<\beta <\alpha$, as shown in \Fref{f:nonfractal}. The resulting lattice has a total number of points that scales as $O(L^{2-\alpha+\beta})$. Choosing $(\alpha-\beta)\geq \frac{1}{2}$ gives the number of sites growing slower than $L^{\frac{3}{2}}$ as required.
		
		Lattices of this type contain as a subgraphs $L^{\alpha}\times L^{\beta}$ polynomial-sized rectangular sections of square lattice. An Ising model on this subgraph would exhibit a phase transition and an exponential memory lifetime, and these properties can be extended to an Ising model on the entire lattice by use of the GKS inequalities described in \Sref{s:gks}. As in the case of the Sierpinski carpet FPCs, quantum product codes based on these graphs would inherit these phase transitions. 
		
		Although this example would not allow for a local realization of the resulting product code in $\R^3$, it may be possible that the additional freedom to use non-fractal graphs such as this might in general allow for more flexibility to construct a local model in $\R^3$.
	
	\subsection{Relation to previous work}	

		Several existing quantum codes have relations to fractal geometry. Two of the most prominent are the Haah codes~\cite{Haah2011} and Yoshida's fractal spin liquids~\cite{Yoshida2013}. While one motivation for considering such codes has also been to engineer self-correcting behaviour, the relationship between our work and these models appears largely superficial. In these previous works, the systems were typically defined as local, translation-invariant Hamiltonians acting on a regular cubic lattice. These systems are engineered so that the support of their logical operators is a fractal subset of the lattice. Significantly, such fractals are of finite ramification. In contrast, the FPC Hamiltonians we study here are themselves defined on a fractal lattice, breaking translation invariance and directly giving rise to the infinitely ramified fractal logical operators. The breaking of translation invariance, along with scale invariance, also directly allows us to escape no-go theorems such as that of Ref.~\cite{Yoshida2011}.
		
		Recently, a new code construction has appeared that produces local subsystem codes with properties inherited from an arbitrary base stabilizer code~\cite{Bacon2014}. Using a concatenated base code, the resulting subsystem code appears to have fractal structure. Again, the relationship between these codes and our work is purely superficial, in particular noting that the FPCs are commuting stabilizer code models while the codes of Ref.~\cite{Bacon2014} are non-commuting subsystem codes.

	\subsection{Numerical simulation}
	
		In the absence of a rigorous proof of self-correction, an attractive strategy is to attempt some numerical simulation of thermalization for an FPC system. However, this could prove prohibitively difficult since the FPCs are only defined for certain (exponentially spaced) system sizes. The smallest FPC family with dimension $<3$, corresponding to $b=14$ and $c=12$, for $l=0,1,2,3$ requires $33$, $3.8\times 10^4$, $8.3\times 10^7$, and $2.1\times 10^{11}$ qubits respectively. Since the $l=0$ case is simply a standard toric code, this leaves very little ability to reasonably simulate these systems. Even by dropping the requirement that the system have dimension $<3$, the smallest code family (with $b=3$, $c=1$) still requires $1.4\times 10^8$ qubits for $l=4$. In order to realistically simulate these systems, it may be necessary to find some way to consistently interpolate between these system sizes. This difficulty may be somewhat alleviated by the use of models based on non-fractal graphs that can be defined on more intermediate lattice sizes as described in \Sref{s:nonfractal}.
	

\acknowledgments{
	We thank Hector Bombin, Ben Brown, Simon Burton, Jeongwan Haah, Steve Flammia, and Pieter Naaijkens for helpful discussions and comments. This work is supported by the ERC grant QFTCMPS and by the cluster of excellence EXC 201 Quantum Engineering and Space-Time Research.
}


\end{document}